%% file: Turbo16ARX.tex
\documentclass[conference,twocolum,final]{IEEEtran}

\usepackage{bbm}
\usepackage{algorithmic}
\usepackage{algorithm} 
\usepackage{amsthm}
\usepackage{amsmath,amssymb}
\usepackage{amsmath}

\usepackage{bbm}

\DeclareMathOperator{\E}{\mathbb{E}}
\DeclareMathOperator{\Prb}{\mathbb{P}}

\newenvironment{compactlist}{
 \begin{list}{{$\bullet$}}{
  \setlength\itemsep{1em}
  \setlength{\itemindent}{\leftmargin}
  \setlength{\leftmargin}{0pt}
 }
}{
 \end{list}
}

\usepackage{subfigure}
\ifCLASSINFOpdf 
\usepackage[pdftex]{graphicx} 
\usepackage[pdftex,colorlinks,
           bookmarks=true,%
            pdftitle=Performance\ of\ spatial\ Multi-LRU\ caching\ under\ traffic\ with\ temporal\ locality,%
           pdfauthor=A.\ Avranas\ -\ A.\ Giovanidis]{hyperref}  
\usepackage[numbers,sort&compress]{natbib} 
\usepackage{hypernat} 
\else

\usepackage[dvips]{graphicx} 
\usepackage[dvips,colorlinks,
           bookmarks=true,%
           pdftitle=Performance\ of\ spatial\ Multi-LRU\ caching\ under\ traffic\ with\ temporal\ locality,%
           pdfauthor=A.\ Avranas\ -\ A.\ Giovanidis]{hyperref}  
\usepackage[numbers,sort&compress]{natbib} 
\fi 
\hypersetup{
   bookmarksnumbered,
   pdfstartview={FitH},
   citecolor={blue},
   linkcolor={red},
   urlcolor=[rgb]{0,0.55,0},
   pdfpagemode={UseOutlines}}

\newtheorem*{proof*}{Proof}

\usepackage[scriptsize,it]{caption}   
\begin{document}

\title{Performance of spatial Multi-LRU caching under traffic with temporal locality}

\author{Apostolos~Avranas$^\dagger$ and Anastasios~Giovanidis$^\ast$\\[2ex]}

\maketitle

\begin{abstract}
In this work a novel family of decentralised caching policies for wireless networks is introduced, referred to as spatial \textit{multi-LRU}. These improve cache-hit probability by exploiting multi-coverage. Two variations are proposed, the \textit{multi-LRU-One} and \textit{-All}, which differ in the number of replicas inserted in the covering edge-caches. The evaluation is done under spatial traffic that exhibits temporal locality, with varying content catalogue and dependent demands. The performance metric is hit probability and the policies are compared to (1) the single-LRU and (2) an upper bound for all centralised policies with periodic popularity updates. Numerical results show the multi-LRU policies outperform both comparison policies. The reason is their passive adaptability to popularity changes. Between the -One and -All variation, which one is preferable strongly depends on the available storage space and on traffic characteristics. The performance also depends on the popularity shape.
\end{abstract}

\begin{keywords}
Wireless; Cache; LRU; Information Centric Networking; Multi-Coverage; Hit Probability; Popularity
\end{keywords}

\let\thefootnote\relax\footnotetext{\hspace{-2ex}$^\dagger$Mathematical and Algorithmic Sciences Lab, France Research Center, Huawei Technologies Co. Ltd., 
Arcs de Seine Bât. A, 20 quai du Point du Jour 92100 Boulogne Billancourt, France; apostolos.avranas@huawei.com\\ 
$^\ast$Universit\'e Paris-Saclay, CNRS \& T\'el\'ecom ParisTech, LTCI lab,
23 avenue d'Italie,
75013 Paris, France; anastasios.giovanidis@telecom-paristech.fr}
\newcommand{\thefootnote}{\arabic{footnote}}

\input{./p1_introduction.tex}

\input{./p2_policy_network.tex}
\input{./p3_temporal_model.tex}

\input{./p4_numerical_eval.tex}

\section{Conclusions}
\label{secVI}
In this work, the proposed Multi-LRU policies, that update each cache content in a per-demand basis, are evaluated under traffic with temporal locality. Their hit probability is compared to the single-LRU, as well as the family of centralised policies with periodic cache updates based on popularity estimates. It is shown that Multi-LRU outperforms single-LRU as well as the centralised policies for a large range of parameter values. The reason is the fast passive adaptability of the policies to traffic changes. The performance strongly depends on coverage and traffic parameters, and notably the content popularity shape.

\input{./p5_appendix.tex}


%
\bibliographystyle{unsrt}
\footnotesize

\end{document}

%% file: p1_introduction.tex
\section{introduction}

The design of today's and future networks is characterised by a paradigm shift, from a host-centric communication
architecture, towards an Information Centric Networking (ICN) one. 
Following this novel concept, network nodes are equipped with storage capacity where data objects can be temporarily cached and retrieved \cite{Paschos1602}. In this way information can be made available closer to the user, it can be accessed reliably \cite{Timo15} with minimum delay, and possibly with a quality adaptable to the users' preferences, as envisioned in the case
of multimedia files. The principal benefit is the reduction of traffic 
flow at the core network by serving demands from intermediate nodes \cite{Sourl11}. 
The edge-nodes constitute a very important part of the ICN architecture, since it is where the wireless users directly have access
to. When these nodes are equipped with storage capability, download path length is minimised \cite{ElRobSIGCOMM15}. 

In this work, we consider the wireless edge of a content centric network, which consists of a set of transmitting nodes taking
fixed positions on a planar area, and a set of users dynamically arriving at this area and asking for service. The set
of transmitters can refer to base stations (BSs) of a cellular network, small stations of heterogeneous networks, WIFI
hotspots, or any other type of wireless nodes that can provide access to users. 
A user can be covered by multiple of these nodes, but she/he will be served by only one.

An important question is how to best manage the available edge-memories, in order to maximise the hit probability of user-demands. We define the hit probability as the probability that a user will find her/his demand cached in
the memory of one of the cells she/he is covered from. By managing, we mean to choose a policy that decides which
objects to install in each cache and how each cache inventory is updated over time. 
For the edge-network with storage capability, there exists a considerable number of policies proposed in the recent literature. To find the optimal content placement that minimises downlink delay, Golrezaei et al \cite{GolrezaeiINFOCOM12} formulate
a binary optimisation problem and
propose approximation and greedy algorithms for its solution. B{\l}aszczyszyn and Giovanidis \cite{BlaGioICC15} provide
a randomised strategy that maximises the hit probability using reduced knowledge over the network topology, i.e. just the 
coverage probability. Poularakis et al \cite{PoulTCOM14} formulate a joint content placement and routing 
problem that maximises the fraction of requests served locally by the deployed small BSs and
propose an approximation algorithm for its solution. Other proposals can be found in \cite{DehgINFO15}, \cite{MassoulieCache15}.

In the above works, the proposed caching policies are centralised and result as solutions of optimisation problems. These use as input considerable system information, including exact knowledge over an assumed static catalogue of files and its popularities, as well as the network topology. Since such knowledge is not trivial to be obtained, while centralised optimisation is not very practical, Giovanidis and Avranas have proposed in \cite{GioAvraSIG16} a novel family of distributed caching policies that profit from multi-coverage, without the need of the above information. These policies are named \textit{spatial multi-LRU} and constitute an extension of the standard single-cache Least Recently Used (single-LRU) principle, to the case with several inter-dependent caches. The multi-LRU policies were evaluated in \cite{GioAvraSIG16} assuming time-stationary traffic of demands, as that in  \cite{GolrezaeiINFOCOM12}, \cite{BlaGioICC15}, \cite{PoulTCOM14}. The Independent Reference Model (IRM) \cite{FaPrSIAM78} describes such traffic, based on which (a) there is a fixed catalogue of requested objects, (b) the popularity of each object is known and constant over time and, (c) every request is independent of previous ones. With IRM input, the authors of \cite{GioAvraSIG16} derive approximate Che-like analytical formulas for the hit probability. Evaluation of the novel policies shows that their performance, without use of excess system information, approaches that of centralised ones.

Although the IRM offers tractability, it is not enough to describe real traffic aspects. In real networks new objects (never requested before) keep on appearing, while older ones become obsolete after some time. Furthermore, the popularity of a content does not remain constant but varies over time, and there is dependence between requests of the same object within some time horizon. All these characteristics are described by the term 
\textit{temporal locality} \cite{BeMASCOTS00}, \cite{TraversoTranMult15}, \cite{OlmosTEMPO14}. Generators of such traffic have been proposed in the literature \cite{Almeida96}. A so-called Shot Noise Model (SNM) is presented in \cite{TraversoTranMult15}, \cite{LeoINFO15}, which we make use here as a basis of our own traffic model. Under SNM the demand process is a superposition of independent Poisson processes (not necessarily homogeneous),
one for each content.

Altogether, in this paper we study the behaviour of spatial multi-LRU policies under a traffic model with temporal locality. We evaluate their performance for hit probability metrics. The importance of this study lies in the significant performance gains exhibited for the novel policies under such traffic with evolving catalogue and varying file popularities. These gains are impressive not only when compared to the \textit{single-LRU} policy, but also to the family of static policies that periodically need to track popularity changes and appropriately update their catalogue. Since, the newly proposed policies do not use any a-priori knowledge (exact or estimated) of traffic-related information, they can adapt to any type of traffic input and (when multi-coverage is available) they always achieve considerable benefits, something not possible under the same input from static/semi-static policies (\cite{GolrezaeiINFOCOM12}, \cite{BlaGioICC15}, \cite{PoulTCOM14}, \cite{DehgINFO15}, \cite{MassoulieCache15}). 

In Section \ref{secII} the spatial multi-LRU policies are introduced. Section \ref{secIII} describes the wireless multi-coverage model. A significant part of the work is dedicated to the generation of spatial traffic with temporal locality and its characteristics (Section \ref{secIV}). The numerical evaluation is given in Section \ref{secV}, where a comparison with the single-LRU, and also with semi-static policies that periodically update their catalogue is provided. Final conclusions are drawn in Section \ref{secVI}. Calculations not included in the main body of the paper can be found in the Appendix.


%% file: p2_policy_network.tex
\section{Spatial multi-LRU}
\label{secII}

In the \textit{single-LRU} policy each user can be connected to a single station (the one with the strongest signal) and can have access to its cache. For a given cache, we index the files placed in the inventory. The first position is called \textit{Most Recently Used (MRU)}, and the last \textit{Least Recently Used (LRU)}. A new request for an object triggers one out of two options. (a. Update) If the object is already in the cache it is moved to the 
MRU position and is downloaded immediately to the user. (b. Insertion) If the object is not in the cache it is downloaded from 
the core network and inserted as new at the MRU position, while the object in the LRU position is evicted.

The multi-LRU policies make use of multi-coverage in the following way. A user can check all caches of covering stations for the demanded object and can download
it from anyone that has it in its inventory. Hence, they exploit the fact that by searching, the user indirectly informs each covering station over the inventory content of its neighbours. Then, cache updates and object insertions can be done in an efficient way for hit performance. Let $\mathcal{M}$, with cardinality $m$, be
the set of caches (belonging to transmitters) that cover the requesting user. We introduce the following policy variations. 

\begin{compactlist}
\item \textbf{multi-LRU-One:} Action is taken only in one of the $m$ caches. (a. Update) If  $\mathcal{M}$
is not empty and the content is found in a subset of $\mathcal{M}$, only one cache from the subset is used for download and,
for this, the content is moved to the MRU position. (b. Insertion) If the object is not found in any of the $m$ caches, it is inserted in only one cache from $\mathcal{M}$ and its LRU object is evicted. This one cache can be chosen as the closest to the user,
a random one, or by some other criterion. In this work we choose the closest node.
\item \textbf{multi-LRU-All:} (a. Update) If the content is found in a non-empty subset of $\mathcal{M}$, all caches
of that subset are updated. (b. Insertion) If the object is not found in any cache of $\mathcal{M}$ the object is 
inserted in all $m$.
\end{compactlist}


\section{Wireless multi-Coverage}
\label{secIII}

We consider the following network model. The transmitter positions form a  2-dimensional square lattice (grid). To every station we allocate storage space which can contain at most $K$ objects (irrespective of the sizes of different files. Equivalently for the model, unit size per object is assumed). Each transmitter has a possibly random area  of wireless coverage associated such that, when users arrive within this area they can be served by the transmitter.
Neighbouring coverage areas can overlap. Hence, a user arriving at a random location may be covered 
by multiple BSs, or may not be covered at all. Let $p_m$ be the
probability of a randomly located user to be covered by $m$ stations. Of course it holds $\sum_{m=0}^{\infty}p_m=1$. The expected number of covering stations
is denoted by $\overline{N_{bs}} :=\sum_{m=0}^{\infty}mp_m$.

Different coverage models define differently the shapes of coverage cells and as a consequence give different values for the coverage probabilities $p_m$. Specific models to be considered are the $\mathrm{SINR}$ model, and the $\mathrm{SNR}$/Boolean model \cite{BartekCOV}. 
For both, the coverage cell of a station is the set of planar points for which the
received signal quality exceeds some given signal-quality threshold $T$. The difference between them is that the $\mathrm{SINR}$ model
refers to networks with interference (e.g. when BSs serve on the same OFDMA frequency sub-slot), whereas the $\mathrm{SNR}$
model, to networks that are noise-limited (e.g. by use of frequency reuse, neighbouring stations do not operate on the
same bandwidth). For the Boolean model the covering cell is a disk having center its station and radius $R_b$. 
It coincides with the $\mathrm{SNR}$ model, when no randomness of signal fading over the wireless channel 
is considered.

%% file: p3_temporal_model.tex
\section{Spatial traffic with temporal locality}
\label{secIV}
In what follows we give a detailed description of the traffic model. A central attribute of real traffic is the temporal evolution of the content catalogue. New objects are constantly born while others become obsolete. We denote by $\mathcal{C}_t$ the catalogue (set) of active objects at time $t$, with cardinality $C(t) := |\mathcal{C}_t|$. The evolution of the catalogue size is a random process. We assume that the arrival of a new object $c_i$ coincides with the time of its first request $t_{i}$. The time instants of first requests (arrivals) are modelled as a homogeneous Poisson 
Point Process (PPP) $\Phi$ on $\mathbb{R}$ with intensity $\lambda_c>0$ [$\frac{objects}{unit-time}$]. Here we take as unit-time $1\ day$. 

A pair of variables is related to each content as an independent mark on the arrival process: (a) The first random variable denoted by $T_{i}$ is the $i$'th content's \textit{lifespan}, which gives the length of time 
period during which it is requested by users, and after the period's end it becomes obsolete. We could allow for the realisation $\tau_{i}$ to take infinite values but in such option the size of the catalogue would grow indefinitely, unless the popularity of different objects tends to zero fast enough. To avoid treating such question for now, we let  
$\tau_{i}<\infty$. With such option the catalogue size $C(t)$ fluctuates over time and remains finite. The time interval of an object is $\Delta t_i:=[t_{i}, t_{i}+\tau_{i})$. (b) The second random variable attached to the object $c_i$ is the \textit{volume} $V_{i}$  i.e. the total number of requests during its lifespan. The pair of values $(\tau_i,v_{i})$ per object is chosen independently of other objects and in the general case should be drawn from a joint probability distribution with a given density $f_{(T, V)}(\tau,v)$, where $T$ and $V$ are the generic variables. This means that in general the two variables should be dependent, to describe the correlation between the longevity of user interest towards an object and the number of times this object is requested. 

To simplify the traffic model it is assumed here that $T$ and $V$ are random variables independent of each other, i.e. $f_{(T, V)}(\tau,v) = f_T(\tau)f_{V}(v)$. This simplification can help with the calculation of main traffic quantities and can allow for easier manipulation without obvious impact on the performance of the caching policies. Both variables (lifespan and volume) follow a Power-law, i.e both $T$ and $V$ are Pareto distributed. This is in coherence with statistical analysis of traffic measurements in the literature for the lifespan \cite{Cheng13} and the volume $V$ \cite{Newman05}. The Pareto distribution in both cases has parameter $\beta>1$ (for the expected value to be finite), and its p.d.f. is given by (here for $V$) $f_V(v)=\frac{{\beta}V_{\min}^{\beta}}{{v}^{\beta+1}}$. Its expected value depends on the values of $\beta$ and $V_{\min}$ through the expression $\mathbb{E}[V] = \frac{\beta V_{\min}}{\beta-1}$. So, for the traffic generation when one of the values $\beta$ or $\E[V]$ is determined, the other one follows. To guarantee $V\in\mathbb{N}_+$ for the samples, we choose $V_{\min}=0.5$ and we round to get discrete values. Sampling from a Pareto distribution, generates Zipf-like distributed sizes of objects due to the Power-law behaviour.

Having sampled the lifespan and number of requests $(\tau_i,v_i)$ for a specific object arriving at $t_i$, it remains to determine how these requests are positioned within $\Delta t_i$. To include additional attributes of temporal locality in the traffic model, we let requests be distributed according to a finite point process (given $V<\infty$) and more specifically a (non-homogeneous) binomial point process (BPP) $\Psi_i$ on $\mathbb{R}^{v_i-1}$ with density function $g_i(t,t_i,\tau_i)$ over $t$,
\begin{equation}
\label{PsiBi}
 \Psi_i \sim \mathrm{Binomial}(\Delta t_i,v_i-1,g_i(t,t_i,\tau_i)). 
 \end{equation}
 We randomly position only $v_i-1$ requests, because the first request always coincides with the time of content arrival $t_i$. The choice of the $\mathrm{Binomial}$ distribution further implies that requests take position independently of each other. Since $g_i(t,t_i,\tau_i)$ describes how each of the $v_i-1$ requests is distributed within $\Delta t_i$ according to the function's \textit{shape}, the higher the value of $g_i$ for some $t$, the more probable it is that a request will appear at that point. From now on, the triplet $(\Delta t_i,v_i-1,g_i)$ will be referred to as the \textit{popularity profile}, and the probability density function (p.d.f.) $g_i(t,t_i,\tau_i)$ as the \textit{popularity shape} of content $c_i$. The popularity shape is an important aspect of the model. For some $c_i$ it holds
\begin{equation}
\label{eqg}
g_i(t,t_i,\tau_i)=0 \quad \mathrm{for} \quad t\notin\Delta t_i,
\end{equation}
and
\begin{equation} \label{volume_pop}
\int_{t_i}^{ t_i+\tau_i} g_i(t,t_i,\tau_i) dt=1.
\end{equation}
The cumulative distribution function (c.d.f.) is 
\begin{eqnarray}
\label{eqGB}
G_i(s,t_i,\tau_i):=\mathbb{P}(t\leq s\ |t_i,\tau_i)=\int_{-\infty}^{s}g_i(t,t_i,\tau_i) dt,
\end{eqnarray}
with $G_i(t_i,t_i,\tau_i)=0$ and $G_i(t_i+\tau_i,t_i,\tau_i)=1$. The joint p.d.f. of the $v_i-1$ requests is their product, due to independence, $g_i^{(v_i-1)}(x_1,\ldots,x_{v_i-1},t_i,\tau_i) = g(x_1,t_i,\tau_i)\cdots g(x_{v_i-1},t_i,\tau_i)$. 

When the requests follow a homogeneous BPP for some $c_i$ with a given $\Delta t_i$, the shape function is \textit{uniform} and takes the expression $g_i(t,t_i,\tau_i)=\tau_i^{-1}\mathbf{1}_{\left\{t\in\Delta t_i\right\}}$. For further shape options we refer the reader to \cite{Richier14}. In this reference work, for finite volume per object three shapes are proposed, namely (i) the \textit{logistic}, (ii) the \textit{Gompertz}, and (iii) the \textit{negative exponential}\footnote{
By assuming that the lifespan of every content ends when it reaches its $1-\varepsilon$ of its total views $v_i$, $\tau_i$ can be mapped to the curve parameter $\lambda$ in \cite{Richier14}, which parametrises the speed of a content's popularity change. In this paper we chose $\varepsilon=0.02$.}. Applying this to our traffic generator, when a new object arrives it is assigned a shape of index $k$ with probability $a_k$, the exact value of which is a tuneable parameter. 
%

The spatial (geographical) aspect of traffic plays an important role in influencing the performance of the policies studied here. In our work requests are uniformly positioned on a finite 2D plane. However, the traffic model can be easily extended to incorporate spatial locality e.g. by enforcing per content requests to follow a spatial 2-dimensional Gaussian distribution around some specified center.

\subsection{Characteristic Quantities of Traffic}
Based on the above description, characteristic quantities of the generated traffic can be derived (see the Appendix).

$\bullet$ \textit{Mean Catalogue Size $\E[C]$.} Because of the stationarity of the arrival PPP
the expected number of active contents (hence catalogue size) does not depend on time $s\in\mathbb{R}_+$,
\begin{eqnarray}
\label{EC}
\E[C] = \E[N_{act}(s)]  & = &  \lambda_c\Prb(V>1)\E[T].
\end{eqnarray}

$\bullet$ \textit{Mean Total Number of Requests within $[0,s]$} [$days$],
\begin{eqnarray}
\label{Nreq}
N_{req}([0,s]) & = & s\lambda_c\E[V].
\end{eqnarray}

$\bullet$ \textit{Cache-to-(Mean)-Catalogue-Size-Ratio (CCSR)} where we omit $\Prb(V>1)$, which is just a scaling constant,
\begin{eqnarray}
\label{CCSR}
\rho & := & \frac{K}{ \lambda_c\E[T]}.
\end{eqnarray}

%% file: p4_numerical_eval.tex
\section{Numerical evaluation}
\label{secV}

\begin{figure*}[th!]    
\centering  
\label{CacheEval2}        
        \subfigure[Hit Prob. VS $\overline{N_{bs}}$ (various cache-size $K$).]{          
           \includegraphics[width=0.315\textwidth]{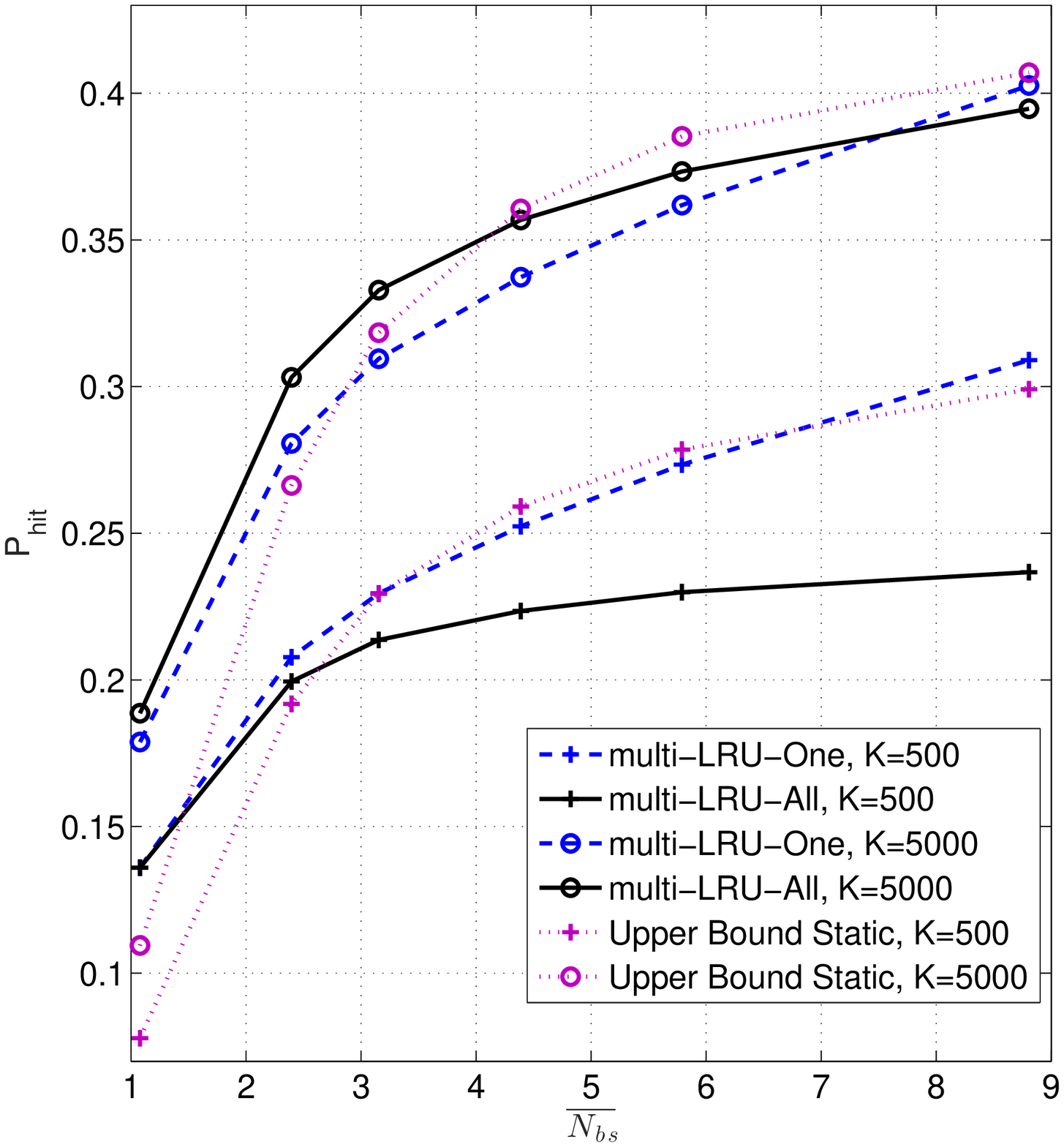}
           \label{fig:Hit_Nbs}
           }
	   \subfigure[Hit Prob. VS CCSR $\rho$ (various $\E{[V]}$).]{          
           \includegraphics[width=0.315\textwidth]{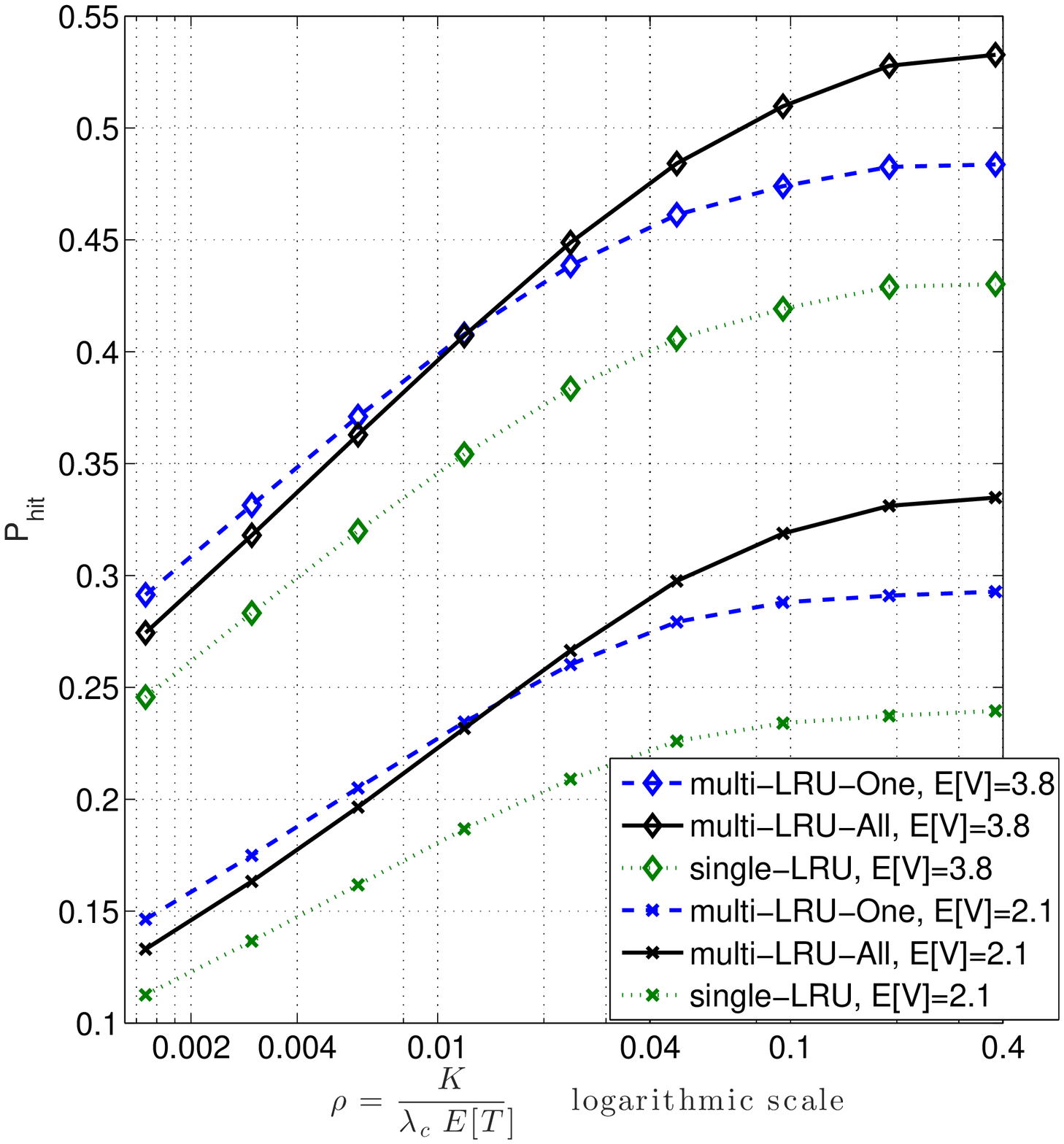}
           \label{fig:Hit_K}
           }
	   \subfigure[Hit Prob. VS $\overline{N_{bs}}$ (different popularity shapes).]{          
           \includegraphics[width=0.315\textwidth]{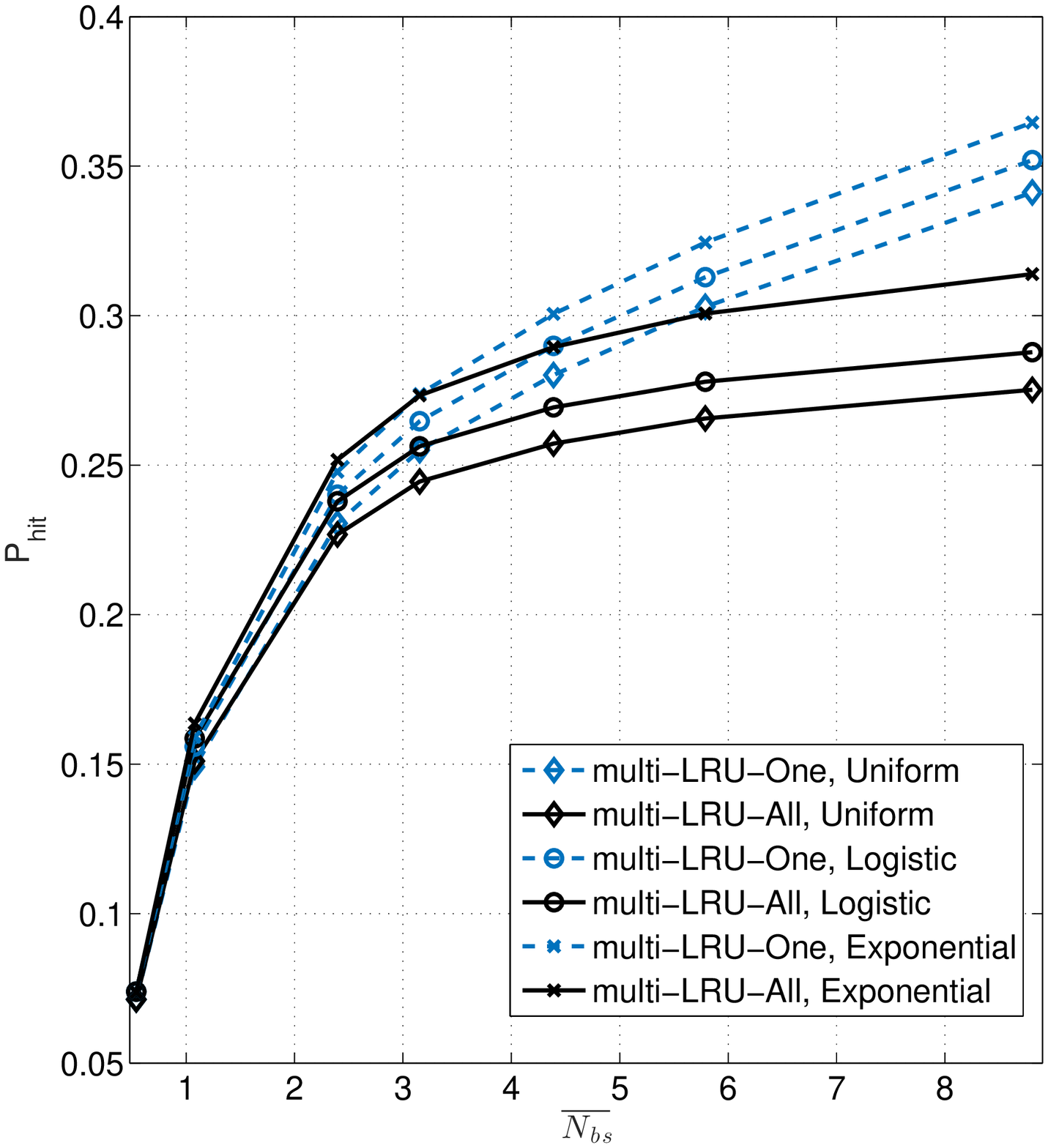}
           \label{fig:Hit_shapes}
           }
\caption{Evaluation of the hit probability of multi-LRU policies for different system variables. Parameter values in (b) $\overline{N_{bs}}=2.4$. In (c) $K=1500$.}
\end{figure*}
For the performance evaluation of the multi-LRU policies extensive simulations are performed. We consider a rectangular
window where $N_{st} = 20$ stations are placed on a square lattice configuration. Assuming a Boolean coverage model every station can serve users within distance $R_b$ from itself. The larger the $R_b$, the bigger the coverage area of each station, and the stronger the multi-coverage effects. Obviously $R_b$ determines the expected number of stations covering a random user, i.e.  $\overline{N_{bs}}$. In the figures to follow, it is $\overline{N_{bs}}$ shown to vary rather than $R_b$, as the former variable is sensitive to both the choice of BS density and the coverage model characteristics.

If not mentioned otherwise, $\lambda_c=2400$ [objects/day], with expected request volume per content $\E[V]=2.1$.  The distribution lifespan p.d.f. is truncated Pareto, with $T\in[\tau_{min},\tau_{max}]= [0.1, 96]$ [$days$], and $\E[T] = 35$. The overall duration of the simulations is $15$ [$months$]. The popularity shape $g_i$ of content $c_i$ is chosen with probability $a_1=0.06$ as Logistic, $a_2=0.38$ as Gompertz and with $a_3=0.56$ as Exponential. Later we include the Uniform shape.

In each simulation, we keep some variables fixed and let others vary, to evaluate the policy performance and understand their influence. The variables to be varied are categorised into (1) \textit{network} and (2) \textit{traffic} variables. The network variables are the mean coverage number $\overline{N_{bs}}$ and the cache size
$K$. The traffic variables are the average request volume per content $\E[V]$, the mean lifespan $\E[T]$ and the probability vector for the shape $(a_1, a_2, a_3)$.

\subsubsection{Network influence}


The studied system has both spatial and temporal dimensions. For a hit to happen, two consecutive requests of the same content should not only be close in time, but should also occur in the coverage area of the same station. So, to increase the hit probability, data-objects should both: stay long-enough in each cache, and be inserted in as many caches as possible. But since storage space per station is limited, a trade-off arises. The multi-LRU-All which for every missed request inserts the object in
all user-covering stations, spreads the object geographically wider at the cost of occupying several memory slots. For each object inserted in a station another one is evicted. As a result, in the -All variation contents stay stored for a shorter time period. On the other hand, multi-LRU-One does not spread the object to neighbouring caches, but the objects stay longer in each memory after insertion. 

This trade-off between geographical expansion and preservation of a content in caches is depicted in Fig. \ref{fig:Hit_Nbs} 
for different values of the memory size. The larger the storage space, the longer it takes for an object to be evicted. So, for large memory the geographical expansion of an object is beneficial. This is shown for $K=5000$ in which case multi-LRU-All surpasses multi-LRU-One. But for each case $K$ there is a value of $\overline{N_{bs}}$ after which the performance of the -All is less than -One, because after this value further increase of content diversity is at the cost of content variety. The smaller the cache size, the more valuable storage space becomes because an object stays less time in the cache before eviction. Hence, for smaller $K$ -One shows better performance and exceeds the -All variation even for small values of multi-coverage $\overline{N_{bs}}$.

In Fig. \ref{fig:Hit_K}  the performance of multi-LRU policies versus the CCSR ratio $\rho$ is illustrated. This ratio is equal to the mean number of memory slots per active content and is a measure of the system's storage capability, because the smaller it is than one, the less storage resources are available. Of course if it is close or larger than one, the average catalogue size can be cached entirely in each station, and hence the hit probability should be close to the "cacheability" limit (the hit probability for all objects that appear at least twice).
Keeping the denominator of $\rho$ constant, Fig. \ref{fig:Hit_K} shows the impact of the memory size
on the policy performance. Obviously, hit probability increases with $K$, but for smaller $K$, the -One variation is preferable to the -All, as explained also previously. There is again a critical point in $K$ after which the -All variation is preferable (for large storage). If the ratio $\rho$ is further increased, the performance gains are diminishing for both variations, and saturation occurs. 

\subsubsection{Traffic influence}

The qualitative impact of the mean lifespan value on the hit probability can be understood by reading Fig. \ref{fig:Hit_K} in the opposite direction of the x-axis, from right to left. Keeping the numerator constant, as $\rho$ decreases $\E[T]$ increases. This means that the same storage capacity serves a larger active catalogue size $\mathbb{E}[C]$. Consequently the overall performance drops. Moreover, Fig. \ref{fig:Hit_K} illustrates that the hit probability improves as the average number of requests per content  $\E[V]$ increases. The reason is that for higher $\mathbb{E}[V]$ a content put in storage is requested and hit more times.

In Fig. \ref{fig:Hit_shapes} each curve corresponds to a scenario where contents are assumed to follow only one particular popularity shape. Specifically, either the logistic, or the negative exponential, or the uniform shape is used. In the negative exponential shape popularity takes big values in a short time period and then drops abruptly. A steep popularity shape implies that consecutive requests of the same content appear close to each other in time. This makes more probable the event that the content is not evicted before its next request happens. With this in mind, it can be understood that the negative exponential can lead to higher hit probabilities than the uniform shape. Interestingly, for isolated caches the authors in \cite{TraversoTranMult15}, \cite{OlmosTEMPO14}  state that the shape does not affect significantly the hit probability of LRU. In our model, this can be observed when $\overline{N_{bs}}$ takes small values so every user can connect to at most one station. This observation is confirmed in Fig \ref{fig:Hit_shapes}. But as $\overline{N_{bs}}$ increases and multi-coverage effects appear, \textit{the multi-LRU performance depends considerably on the correlation between requests of the same content, and thus the shape of the popularity curves}.

\subsubsection{Comparison with the single-LRU}

Under single-LRU a user can access only one (the closest in this work) station's memory even when covered by more than one. As a result hit performance is independent of $\overline{N_{bs}}$ (provided coverage is enough so that a user is always covered by at least one station). Depriving the user of the ability to retrieve its content from all covering stations, strongly reduces the overall hit probability. That's why in Fig. \ref{fig:Hit_K} where $\overline{N_{bs}}=2.4$ is relatively small, both multi-LRU policies show relative gains compared to the single-LRU, for every value of $\rho$. The maximum gains reach 30\% when $\E[V]=2.1$ and 20\% when $\E[V]=3.8$.

\subsubsection{Comparison with centralised Policies with periodic Popularity updates and prefetching (POP)}

As mentioned in the introduction, most caching policies proposed in the literature distribute contents to memories in a centralised way \cite{GolrezaeiINFOCOM12}, \cite{BlaGioICC15}. These caching algorithms (POP) use as input the popularity of the contents, assumed known and constant over time. But, when the traffic exhibits temporal locality, popularities change perpetually and should be estimated regularly. Based on every new estimation, memories should be updated. Specifically we can assume that at time $t_n=n\Delta t_{ev}$, $n\in\mathbb{Z}$, the caches are updated by the POP using the estimated popularities during the time interval $[t_n-\Delta t_{pop},t_n)$. Let $\mathcal{F}_{x,t_n}$ be the set of the $x$ most requested contents in $[t_n-\Delta t_{pop},t_n)$. 

\textit{We propose an upper bound}. Let a user requesting content $c$ connect to $m$ stations with probability $p_m$. Then the user "sees" $mK$ memory slots. The most favorable scenario is when all memory slots are filled at time $t_n$ by the $mK$ most popular files until that time, i.e. the content set $\mathcal{F}_{mK,t_n}$. Then the maximum hit probability within the time interval $[t_n,t_n+\Delta t_{ev})$ is 
\begin{eqnarray}
\label{POPub}
P_{hit}^{(POP)}[t_n,t_n+\Delta t_{ev})\leq\sum_{m=1}^{\infty}p_m\Prb(c\in\mathcal{F}_{mK,t_n}).
\end{eqnarray}
Clearly this upper bound depends on the intervals $\Delta t_{ev}$ and $\Delta t_{pop}$. The more regularly the algorithm updates the caches the better performance it achieves. But considering that a caching update will use backhaul and computational resources by the controller, $\Delta t_{ev}$ cannot be too small. We can assume that $\Delta t_{ev}=1$ $day$ and the caching policy runs every night when the request load is low. As far as $\Delta t_{pop}$ is concerned there is a "crisp" optimal choice. If it is too big, $\mathcal{F}_{mK,t_n}$ will possibly include outdated contents. On the other hand, small $\Delta t_{pop}$ can result in excluding even the popular objects from $\mathcal{F}_{mK,t_n}$ because they have not been sufficiently requested. The bound in (\ref{POPub}) was evaluated by Monte Carlo simulations. The optimum $\Delta t_{pop}$ number of days was found by increasing it until $P_{hit}^{(POP)}$ starts decreasing. Fig. \ref{fig:Hit_Nbs} depicts $P_{hit}^{(POP)}$ for two memory sizes, where $\Delta t_{pop}$ was chosen (among a large set of possibilities) equal to 5 and 10 days, for $K=500$ and $5000$ respectively. 

Fig. \ref{fig:Hit_Nbs} shows that even the upper bound for POP does not surpass in performance the appropriate multi-LRU policy, except maybe for a small range of $\overline{N_{bs}}$. In \cite{GioAvraSIG16} where the traffic was assumed static (IRM), multi-LRU performed lower than the centralised policies POP. But under a temporal traffic model (which is also more realistic), the ability of multi-LRU policies to update at each request the caches, without the need to estimate the content popularities, results in a considerable performance boost. 

\label{secVI}

%% file: p5_appendix.tex
\appendix{A) \textit{Expected Catalogue Size}.} This is equal to the expected number of active objects at time $s$ $N_{act}(s)$. There are three logical criteria for an object to be active at time $s$. The first one is that its volume is larger than $1$, $\left\{V_i>1\right\}$, because otherwise objects with volume $v_i=1$ will be extinct as soon as they appear, and their lifespan (chosen independently of $v_i$) does not play any role. The second criterion is that it arrives at $\left\{t_i<s\right\}$. The third one is that $\left\{s<t_i+T_i\right\}$. The last two criteria can be described by the joint condition $\left\{s\in\Delta t_i\right\}$.
\begin{eqnarray}  
\E[C] & = & \E[N_{act}(s)]  \nonumber\\
& = & \E\left[ \sum_{i:t_{i}<s}\mathbf{1} _{\lbrace{V_{i}>1}\rbrace}\mathbf{1} _{\lbrace{t_{i}+T_{i}>s}\rbrace} \right]\nonumber \\ 
& \stackrel{Campbell}{=} & \int_{-\infty}^{s}\E[\mathbf{1} _{\lbrace{V_{i}>1}\rbrace}\mathbf{1} _{\lbrace{T_{i}>{s-t}}\rbrace}]  \lambda_c dt\nonumber\\
& \stackrel{iid}{=} & \int_{-\infty}^{s}  \Prb(V>1) \Prb(T>{s-t})\lambda_c dt\nonumber\\
& \stackrel{u=s-t}{=} & \lambda_c \Prb(V>1) \left(- \int_{+\infty}^{0} \Prb(T>{u}) du\right)\nonumber\\
& = & \lambda_c\Prb(V>1)\E[T].
\label{active_c}
\end{eqnarray}

B)  \textit{Expected Total Number of Requests within the time interval $B_s=[0,s]$}. The way we have defined traffic, the total number of requests will be equal to the number of content arrivals in $B_s$, increased by the number of additional requests from any arrival occurring before $s$. (Note $t^+:=\max\left\{0,t\right\}$)
\begin{eqnarray} 
\label{Req1}
& & \E[N_{req}(B_s)] = \E\left[{\sum_{i:t_i<s}\left(\mathbf{1}_{\left\{t_i\in B_s\right\}}+\mathbf{1}_{\left\{V_i>1\right\}}\Psi([t_i^+,s])\right)} \right]\nonumber\\ 
& & = \int_0^s\lambda_c dt + \sum_{v>1}\mathbb{P}(V=v)\int_{-\infty}^s\E\left[\Psi([t^+,s])\ |v\right]\lambda_c dt.\nonumber\\
\end{eqnarray}
The expectation within the integral is the expected number of requests per content within the interval of interest $B_s$, that is 
\begin{eqnarray}
\label{Ereq}
\E\left[\Psi([t^+,s])\ |v\right] & = &\int_0^{\infty}f_T(\tau) \E\left[\sum_{j=2}^v\mathbf{1}_{\left\{0<x_j\leq s\right\}} \ |t,\tau\right]d\tau\nonumber\\
& = & (v-1)  \int_0^{\infty}f_T(\tau)\mathbb{P}\left(0<x\leq s\ |t,\tau\right)d\tau.\nonumber\\
\end{eqnarray}
By substitution of (\ref{Ereq}) in (\ref{Req1}) the righthand side integral in (\ref{Req1}) can be rewritten as (we omit here $(v-1)$)
\begin{eqnarray}
\label{sub1}
& & \int_{-\infty}^s\int_0^{\infty}f_T(\tau)\mathbb{P}\left(0<x\leq s\ |t,\tau\right)d\tau \ \lambda_cdt\nonumber\\
& \stackrel{Fubini}{=} & \int_0^{\infty}f_T(\tau)\int_{-\infty}^s\mathbb{P}\left(0<x\leq s\ |t,\tau\right) dt \ \lambda_cd\tau\nonumber\\
& \stackrel{shape}{=} & \int_0^{\infty}f_T(\tau)\int_{-\infty}^s\int_0^s g(x,t,\tau) \lambda_cdx \ dt\ d\tau\nonumber\\
& \stackrel{Fubini, (a)}{=} & \int_0^{\infty}f_T(\tau)\int_0^s \int_{-\infty}^s g(x-t,0,\tau) \lambda_c dt \ dx \ d\tau\nonumber\\
& \stackrel{u=x-t}{=} & \int_0^{\infty}f_T(\tau)\int_0^s \int_{x-s}^{+\infty} g(u,0,\tau) \lambda_c du \ dx \ d\tau\nonumber\\
& \stackrel{(\ref{eqGB})}{=} & \int_0^{\infty}f_T(\tau)\int_0^s \left(1-G(x-s,0,\tau)\right) \lambda_c dx \ d\tau\nonumber\\
& \stackrel{(b)}{=} & \lambda_c s.
\end{eqnarray}
%
In (a) we use the fact that $g(x,t,\tau)=g(x-t,0,\tau)$. In (b) $G(x-s,0,\tau)=0$ for every $x\in[0,s]$ because $x-s\leq 0$, and $t=0$ is the arrival of the content. Finally, combining (\ref{sub1}) multiplied by $(v-1)$ that was omitted, with (\ref{Req1}) we get 
\begin{eqnarray}
\label{Efinal}
\E[N_{req}(B_s)] & = & s\lambda_c \left(1+\sum_{v>1}(v-1)\Prb({V}=v)\right)\nonumber\\ 
&\stackrel{(c)}{=} & s\lambda_c\left(\sum_{v>1}v\Prb({V}=v)+\Prb({V}=1)\right)\nonumber\\ 
&= & s\lambda_c\E[V], 
\end{eqnarray}
where in (c) the equality $1=\sum_{v\geq1}\Prb({V}=v)$ was used. The result is intuitive and reasonable. It can also be extended to the general case where $V$ and $T$ are dependent. Furthermore, the same result is derived if the 
objects have different popularity shape $g_n(t,t_{i},\tau_{i})$ with a  probability $a_n$.